\documentclass[journal,10pt]{IEEEtran}
\ifCLASSINFOpdf

\else

\fi

\usepackage[utf8]{inputenc}
\usepackage{amssymb}
\usepackage{stfloats}
\usepackage{placeins}
\usepackage{overpic}
\usepackage[font=small]{caption}
\usepackage{amsmath}
\usepackage{comment}
\usepackage{mathtools}
\usepackage{booktabs}
\DeclarePairedDelimiter{\norm}{\lVert}{\rVert}
\usepackage{cite}
\usepackage{subfig}
\usepackage{lipsum}
\usepackage{graphicx}
\usepackage{multicol}
\usepackage{algorithm}
\usepackage[noend]{algpseudocode}
\usepackage{tikz}
\usepackage[english]{babel}
\usepackage{amsthm}
\theoremstyle{plain}
\usepackage{dirtytalk}

\graphicspath{ {./images/} }
\usepackage{amsmath}

\usepackage{mathtools}

\usepackage{tabularx}
\newcolumntype{L}[1]{>{\raggedright\arraybackslash}p{#1}}
\newcolumntype{C}[1]{>{\centering\arraybackslash}p{#1}}
\newcolumntype{R}[1]{>{\raggedleft\arraybackslash}p{#1}}
\usepackage{multirow}
\usepackage{mathtools, cuted}

\begin{document}


\title{Rate-Splitting Multiple Access for 6G -- Part I: Principles, Applications and Future Works}


\author{\IEEEauthorblockN{Anup~Mishra, \IEEEmembership{Graduate Student Member, IEEE}, Yijie~Mao, \IEEEmembership{Member, IEEE},  Onur~Dizdar, \IEEEmembership{Member, IEEE} \\and Bruno~Clerckx, \IEEEmembership{Fellow, IEEE}\vspace{0.2cm}\\\textit{(Invited Paper)}\vspace{-0.6cm}}
\thanks{The authors Anup Mishra, Onur Dizdar and Bruno Clerckx are with the Department of Electrical and Electronic Engineering, Imperial College London, London SW7 2AZ,
U.K. (e-mail: anup.mishra17@imperial.ac.uk; o.dizdar@imperial.ac.uk; b.clerckx@imperial.ac.uk).}
\thanks{Yijie Mao is with the School of Information Science and Technology, ShanghaiTech University, Shanghai 201210, China (e-mail: maoyj@shanghaitech.edu.cn).}}


\maketitle


\begin{abstract}
This letter is the first part of a three-part tutorial focusing on rate-splitting multiple access (RSMA) for $\mathbf{6}$G. As Part I of the tutorial, the letter presents the basics of RSMA and its applications in light of $\mathbf{6}$G. To begin with, we first delineate the design principle and basic transmission frameworks of downlink and uplink RSMA. We then illustrate the applications of RSMA for addressing the challenges of various potential enabling technologies and use cases, consequently making it a promising next generation multiple access (NGMA) scheme for future networks such as $\mathbf{6}$G and beyond. We briefly discuss the challenges of RSMA and conclude the letter. In continuation of Part I, we will focus on the interplay of RSMA with integrated sensing and communication, and reconfigurable intelligent surfaces, respectively in Part II and Part III of this tutorial.    
\end{abstract}



\vspace{-0.1cm}
\section{Introduction}\label{Intro}


\IEEEPARstart{T}{}
he advent of $6$G is expected to provide highly reliable and ubiquitous wireless connectivity with significantly improved key performance indicators compared to those in existing communication standards. Specifically, {the enabling technologies of $6$G} are required to achieve $50$ times higher peak data rate, $10$ times reduced latency, and $10^{4}$ times higher reliability (packet error rate) than $5$G\cite{6G@Beyond}. Moreover, core services such as enhanced mobile broadband (eMBB), ultra reliable low latency communications (URLLC) and massive machine type communications (mMTC) or combinations thereof, and emerging services with converging functionalities such as communication, sensing, localization and computing  will address the requirements of use cases which need massive access to cater societal applications ranging from healthcare to infrastructure and industry automation\cite{Onur@6G}. The enabling technologies and potential use cases present a unique set of challenges for $6$G. One such formidable challenge is the issue of designing sophisticated \textit{multiple access (MA)} {schemes\cite{mao2022ratesplitting}}. {MA refers to the broad set of techniques that enable to have multiple users, services (eMBB, URLLC, mMTC, etc), and wireless capabilities (communications, sensing, localization, etc) to share the same resources}. To ensure massive access and provide high quality of service (QoS), it is imperative to depart from the conventional MA schemes and pursue a transmission strategy that measures up to the peculiar challenges of $6$G.
\par Previous generations of cellular wireless networks employed orthogonal MA (OMA) schemes to separate users in the frequency domain ($1$G), time domain ($2$G), code domain ($3$G) or both time and frequency domain using orthogonal frequency division MA (OFDMA) ($4$G and $5$G). The choice of orthogonal radio resource allocation was predominately motivated by the objective of avoiding inter-user interference and high transceiver complexity\cite{mao2022ratesplitting}. However, such an approach leads to an inefficient use of radio resources. To address this issue, aside OFDMA, $4$G and $5$G have exploited multiple antennas using space division MA (SDMA), such as in multi-user multiple-input multiple-output (MIMO) and massive MIMO (MaMIMO), to serve multiple users in a non-orthogonal manner in the same time-frequency block and separate them in the spatial domain using precoders or spatial beams. In contrast to OMA where multi-user interference is completely avoided, in SDMA, multi-user interference can still occur but precoders/beams are designed such that it remains as much as possible subdued compared to the noise. 
\par Another non-orthogonal strategy that has gained traction in $5$G is non-orthogonal MA (NOMA), which in its widely studied form, i.e., power domain NOMA (PD-NOMA), serves multiple users in the same time-frequency block, and separates them in the power domain in single-antenna systems\cite{Lui@NGMA}. PD-NOMA allows superposition of messages of multiple users for transmission, and the multi-user interference ensuing from this non-orthogonal transmission is decoded and removed using successive interference cancellation (SIC) at the receiver. In single-antenna systems, e.g. single-input single-output (SISO) broadcast channel (BC), PD-NOMA has been shown to achieve higher spectral efficiency (SE) than OMA and simultaneously serve higher number of users at an additional cost of increased transceiver complexity. Unfortunately, the design principle of forcing a user to fully decode the interference from other users limits the advantages of PD-NOMA to single antenna settings in the downlink as it fails to fully exploit the spatial domain and the SIC receivers in multi-antenna settings \cite{Bruno@NOMA}. {In the uplink, PD-NOMA relies on time sharing to achieve the capacity region which induces communication overhead and strict synchronization requirements to coordinate  transmissions among all users\cite{mao2022ratesplitting,uplink@MAC}}. 
\par Since multiple antennas are nowadays unavoidable in modern wireless networks, SDMA combined with OFDMA has been the most popular MA scheme in the past 20 years due to its good performance vs complexity trade-off. SDMA, nevertheless, has some drawbacks. Naturally, the number of users that can be simultaneously served by SDMA is limited by the number of antennas (which is captured through the so-called degree of freedom-DoF\footnote{The DoF can be interpreted as the number of interference free streams that can be simultaneously transmitted per channel use\cite{Bruno@NOMA}.}). Moreover, SDMA is highly sensitive to channel state information at the transmitter (CSIT) inaccuracies and incurs performance  loss (DoF and SE both) in the presence of imperfect CSIT in the downlink. In order to fully exploit both the spatial and power domains, cluster-based NOMA was studied where users are grouped into clusters and users within a cluster share the same precoder. However, sophisticated user grouping and ordering algorithms employed in cluster-based NOMA increase computational complexity and still incur DoF loss\cite{Bruno@NOMA}. In addition, like SDMA, cluster-based NOMA is vulnerable to CSIT impairments which leads to poor management of inter-group interference in the downlink \cite{Bruno@NOMA}. {In the uplink, SDMA relies on SIC (like NOMA) to achieve the corner points of the capacity region and therefore is subjected to the same limitation of time sharing  \cite{mao2022ratesplitting}}.
\par To overcome the limitations of the hitherto MA schemes, a novel transmission strategy named rate-splitting MA (RSMA) based on the concept of rate-splitting (RS) was studied and has been shown to generalize several existing MA techniques\cite{uplink@MAC,mao2017rate}. RSMA is a robust interference management strategy that relies on splitting user messages into multiple parts. The split parts are transmitted by employing superposition coding at the transmitter and are decoded using SIC at the receivers. In the same time-frequency block, RSMA separates the transmitted message parts in both the power domain (as in PD-NOMA) and the spatial domain (as in SDMA), and manages interference by adjusting the message split and power allocated to different parts. This novel approach allows RSMA to fully exploit the available time, frequency, power and spatial domains, achieving the optimal DoF with both perfect and imperfect CSIT in the downlink, and capacity region in the uplink without employing time sharing among users, as opposed to PD-NOMA and SDMA\cite{uplink@MAC,mao2022ratesplitting}. Thanks to its powerful interference management capability, RSMA is more robust to CSI impairments, adaptive to network load, and user deployments in wireless networks\cite{Bruno@NOMA}. 
\par In this letter, we overview the design principle and transmission frameworks of downlink and uplink RSMA in Section \ref{Principles}. The applications and future works of RSMA for different enabling technologies and potential use cases of $6$G are discussed in Section \ref{Applications}. Section \ref{Challenges} briefly discusses the challenges on RSMA and we conclude the letter in Section \ref{Conclusion}. 



\section{RSMA: Design Principle and Transmission Framework}\label{Principles}
{The design principle of RSMA is to partially decode the interference and partially treat the interference as noise, thereby generalizing the two interference management strategies of PD-NOMA and SDMA\cite{mao2022ratesplitting}. Although the design principle of RSMA is same for uplink and downlink, the motivation is different, i.e., to avoid time sharing in the uplink, and to enhance the SE over current MA schemes in the downlink. In the following, the downlink and uplink transmission frameworks of RSMA are provided to explicate how it achieves to manage interference effectively.}
\subsection{RSMA in the downlink}\label{DL_RSMA}
RSMA has been studied in various forms depending on the way users' messages are split, and how the split sub-messages are combined into common messages. For ease of illustration, we confine ourselves to the downlink transmission framework based on 1-layer RS\cite{RSintro16bruno,RS2016hamdi} and refer interested readers to \cite{mao2022ratesplitting} for a comprehensive study on different forms of RSMA. {Henceforth, 1-layer RS will be referred to as ‘‘RSMA'’.}
\par We consider a multiple-input single-output (MISO) BC, where $K$ single-antenna users indexed by the set $\mathcal{K}=\{1,\ldots,K\}$, are simultaneously served by a base-station (BS) equipped with $N_{t}$ transmit antennas. In RSMA, the message of user-$k,\,\forall k\in\mathcal{K}$ denoted as $W_{k}$ is split into a common part $W_{\textrm{c},k}$ and a private part $W_{\textrm{p},k}$. The common parts of all users, $\{W_{\textrm{c},1},\ldots,W_{\textrm{c},K}\}$, are combined together to form a single common message denoted by $W_{\textrm{c}}$ which in turn is encoded into the common stream $s_{\textrm{c}}\in\mathbb {C}$. On the other hand, the private part of the message of user-$k,\, \forall k\in\mathcal{K}$ is independently encoded into the private stream $s_{\textrm{p},k}\in\mathbb{C}$. Linear precoders $\mathbf{p}_{\textrm{c}}\in\mathbb{C}^{M}$ and $\mathbf{p}_{\textrm{p},k}\in\mathbb{C}^{M},\,\forall k\in\mathcal{K}$ are used to precode the common and private streams, respectively. The transmit signal $\mathbf{x}\in\mathbb{C}^{M}$ is expressed as:
\begin{equation}\label{eq:Tx_signal}
    \vspace{-0.15cm}
    \mathbf{x}=\mathbf{p}_{\textrm{c}}s_{\textrm{c}}+\sum_{k=1}^{K}\mathbf{p}_{\textrm{p},k}s_{\textrm{p},k},
    \vspace{-0.10cm}
\end{equation}
where $\mathbf{x}$ is subject to the power constraint $\mathbb{E}\{\norm{\mathbf{x}}^{2}\}\leq P$. Denoting the channel between the BS and user-$k$ as $\mathbf{h}_{k}\in\mathbb{C}^{M}$, the received signal at user-$k$, i.e., $y_{k}\in\mathbb{C}$ is given by:
\begin{equation}\label{eq:Rx_Signal}
\vspace{-0.1cm}
    y_{k}=\mathbf{h}_{k}^{H}\mathbf{x}+z_{k},
\vspace{-0.025cm}
\end{equation}
where $z_{k}\sim\mathcal{CN}(0,\sigma_{z}^{2})$ is the additive white Gaussian noise (AWGN) at user-$k$. At the receiver side, user-$k$ first decodes the common stream by treating all private streams as noise. Hence, the signal-to-interference-plus-noise ratio (SINR) for the common stream at user-$k$ is expressed as:
\begin{equation}\label{eq:SINR_C}
    \gamma_{\textrm{c},k}=\frac{{\lvert \mathbf{h}_{k}^{H}\mathbf{p}_{\textrm{c}}\rvert}^{2}}{\sum_{i=1}^{K}{\lvert \mathbf{h}_{k}^{H}\mathbf{p}_{\textrm{p},i}\rvert}^{2}+\sigma_{z}^{2}}.
\end{equation}
 \begin{figure}[!t]
    \centering
    \includegraphics[width=\columnwidth,height=4.9cm]{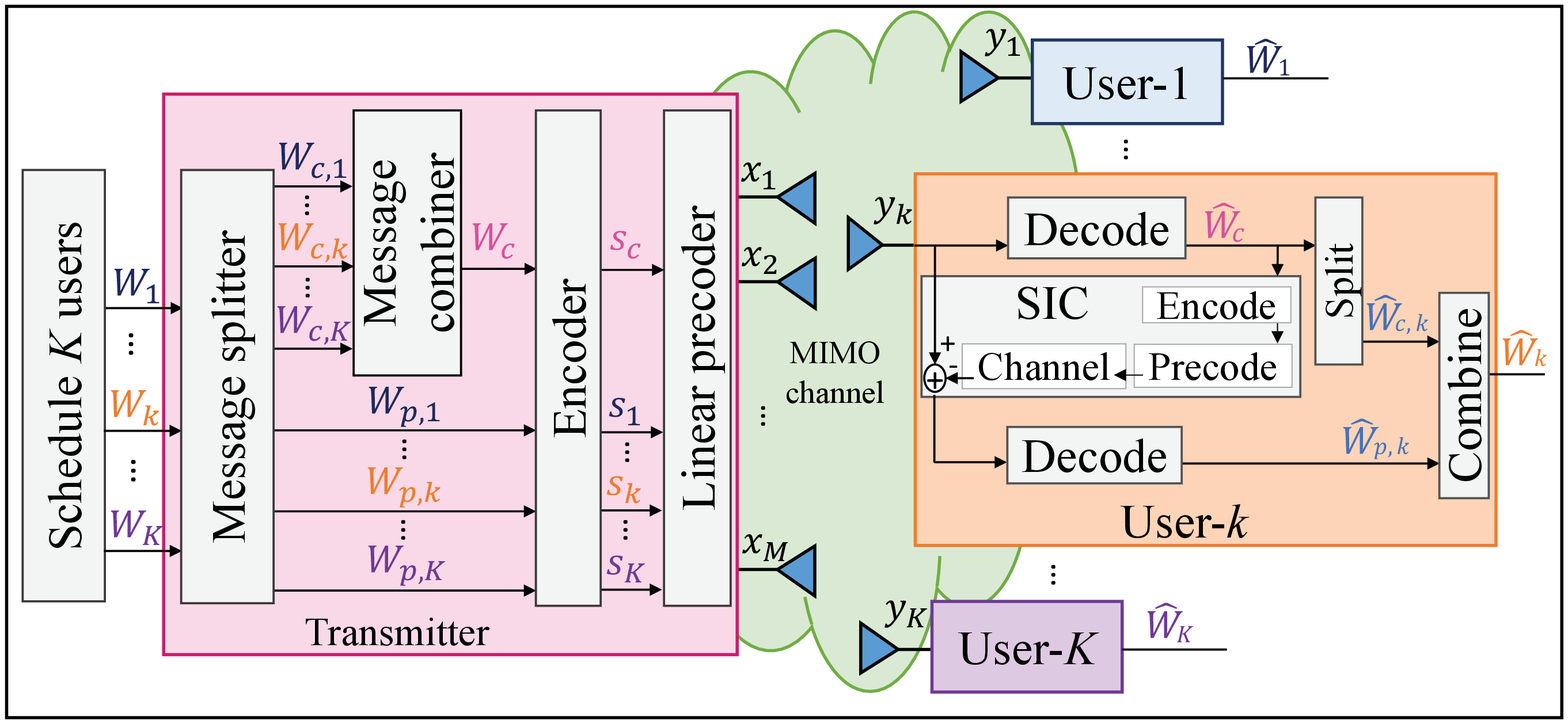}
    \caption{Transceiver architecture of $K$-user downlink RSMA\cite{mao2022ratesplitting}.}%
    \label{fig:RSMA_DL_1LayerRS}\vspace{-0.4cm}
\end{figure}
After successfully decoding and removing the common stream using SIC, user-$k$ decodes its own private stream by treating the private streams of other users as noise. Thus, the SINR for the private stream of user-$k$ is expressed as:
\begin{equation}\label{eq:SINR_P}
    \gamma_{\textrm{p},k}=\frac{{\lvert \mathbf{h}_{k}^{H}\mathbf{p}_{\textrm{p},k}\rvert}^{2}}{\sum_{i=1, i\neq k}^{K}{\lvert \mathbf{h}_{k}^{H}\mathbf{p}_{\textrm{p},i}\rvert}^{2}+\sigma_{z}^{2}}.
\end{equation}
User-$k$ reconstructs its intended message by extracting {$\widehat{W}_{\textrm{c},k}$ from $s_{\textrm{c}}$, extracting $\widehat{W}_{\textrm{p},k}$ from $s_{\textrm{p},k}$, and combining them to form $\widehat{W}_{k}$}. Fig.~\ref{fig:RSMA_DL_1LayerRS} illustrates the downlink transmission framework of RSMA.\footnote{{The transceiver architecture of RSMA in Fig.~\ref{fig:RSMA_DL_1LayerRS} can be extended from MISO to MIMO BC by splitting and encoding a vector of user's messages into a vector of common streams and a vector of private streams\cite{mishra2021ratesplitting,mao2022ratesplitting}}.} Note that the common stream is supposed to be decoded by all users. Therefore, the maximum achievable rate for the common stream is $R_{\textrm{c}}=\log_{2}(1+\gamma_{\textrm{c}})$, where $\gamma_{\textrm{c}}=\min \{\gamma_{\textrm{c},k}\}_{k=1}^{K}$. Moreover, since the common message has common parts of the users' messages, we have {$R_{\textrm{c}}=\sum_{k=1}^{K}C_{k}$}, where $C_{k}$ is the share of the rate of the common stream intended for user-$k$. Therefore, the total rate for user-$k$ is computed as $R_{k}=R_{\textrm{p},k}+C_{k}$, where $R_{\textrm{p},k}=\log_{2}(1+\gamma_{\textrm{p},k})$. Precoders can be designed using low-complexity methods, such as linear zero-forcing precoders for the private streams and multicast precoders (e.g. dominant singular vector of concatenated channel matrix) for the common stream. Alternatively, the precoders can be optimized under different objectives such as maximizing the weighted sum-rate, maximizing the energy efficiency (EE), etc\cite{mao2017rate,mao2022ratesplitting}.
\par {It should be noted that RSMA encapsulates SDMA and is capable of fully utilizing the spatial domain\cite{Bruno@NOMA}}. The presence of the common stream allows RSMA to exploit the power domain and adjust the amount of interference that needs to be decoded by users. Depending on the CSIT accuracy and the precoder design objective, RSMA adjusts the message split ratios and power allocated to the common and private streams at the transmitter. At the user side, user-$k$ decodes part of the interference ($W_{\textrm{c},i},\,\forall i\neq k$) and treats the remaining part as noise ($W_{\textrm{p},i},\,\forall i\neq k$). By this approach, RSMA achieves to softly bridge the two extreme strategies of dealing with multi-user interference, i.e., NOMA which fully decodes interference and SDMA which fully treats interference as noise. The RSMA downlink framework has been rigorously studied from both the information and communication theoretic perspective. RSMA is shown to outperform OMA, NOMA, and SDMA in DoF, SE, and EE performance under different CSIT constraints (perfect and imperfect CSIT) and network loads (underloaded and overloaded)\cite{mao2017rate,Bruno@NOMA}. The performance gains of RSMA are not just limited to the assumption of Gaussian signalling and infinite block lengths but are realized for practical systems as well in throughput performance via link-level simulations\cite{mishra2021ratesplitting}. {Multi-layer RSMA strategies have also been thoroughly investigated and are shown to outperform SDMA and NOMA in SE performance with imperfect CSIT\cite{mao2022ratesplitting,mao2017rate}}.
\vspace{-0.3cm}
\subsection{RSMA in the uplink}
\begin{figure}
    \centering
    \includegraphics[width=\columnwidth]{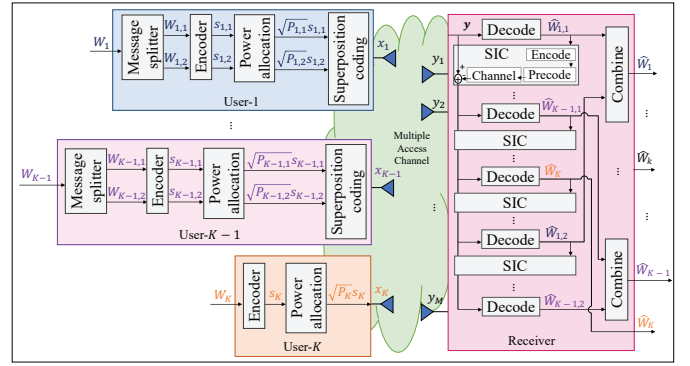}
    \caption{Transceiver architecture of $K$-user uplink RSMA\cite{mao2022ratesplitting}.}%
    \label{fig:RSMA_UL}\vspace{-0.6cm}
\end{figure}

To discuss the uplink transmission framework of RSMA, {we consider the dual multiple access channel (MAC) of the downlink BC in subsection \ref{DL_RSMA}}. {To avoid time sharing and achieve every point of the capacity region in the $K$-user case, splitting the messages of only $K-1$ users is sufficient \cite{uplink@MAC}}. Without loss of generality, we assume that messages of all users except user-$K$ are split. Following, user-$k,\,k\in\mathcal{K}\setminus\{K\}$ splits its message $W_{k}$ into two parts $W_{k,1}$ and $W_{k,2}$. Subsequently, the two parts $W_{k,1}$ and $W_{k,2}$ are independently encoded into streams $s_{k,1}\in\mathbb{C}$ and $s_{k,2}\in\mathbb{C}$, such that $\mathbb{E}\{{\lvert s_{k,i}\rvert}^{2}\}=1,\,i\in\{1,2\}$. The two streams are then respectively allocated powers $P_{k,1}$ and $P_{k,2}$ and superposed together. At user-$K$, the message $W_K$ is directly encoded into $s_K$ and allocated with power $P_{K}$. The transmit signal of user-$k$, $x_{k}\in\mathbb{C}$, is given by:
{
\begin{equation}\label{eq:Tx_UL}
  x_k=\begin{cases}
\sqrt{P_{k,1}}s_{k,1}+\sqrt{P_{k,2}}s_{k,2}, & \textrm{if}\;k\in\mathcal{K}\setminus\{K\}, \\
\sqrt{P_{K}}s_{K}, &  \textrm{if}\;k=K.
\end{cases}
\end{equation}}
The received signal $\mathbf{y}\in\mathbb{C}^{M}$ at the BS is expressed as:
\begin{equation}\label{eq:Rx_UL}
    \mathbf{y}=\sum_{k\in\mathcal{K}}\mathbf{h}_{k}x_{k}+\mathbf{n}_{\textrm{ul}},
\end{equation}
where $\mathbf{n}_{\textrm{ul}}$ is the AWGN noise vector whose elements have zero mean and variance $\sigma_{\textrm{ul}}^{2}$. The BS employs receive filters $\mathbf{d}_{k,1}\in\mathbb{C}^{M}$, $\mathbf{d}_{k,2}\in\mathbb{C}^{M}$ to detect the two streams of user-$k$, $\forall  k\in \mathcal{K}\setminus\{K\}$, and $\mathbf{d}_{K}\in\mathbb{C}^{M}$ to detect the stream of user-$K$. Let $\pi$ denote the decoding order in which the $2K-1$ streams $\{s_{k,i}, s_{K} \mid k\in\mathcal{K}\setminus\{K\}, i\in\{1,2\}\}$ will be decoded such that the stream $s_{k,i}$ will be decoded before stream $s_{k',i'}$ if $\pi_{k,i}<\pi_{k',i'}$. For user-$K$, $\pi_{K,i}$ is simplified to $\pi_{K}$. Following, the rates of decoding streams $s_{k,i}$ and $s_{K}$ are given by:
\begin{equation}\label{eq:UL_Rate}
    R_{k,i}=\log_{2}\bigg(1+\frac{P_{k,i}{\lvert \mathbf{d}_{k,i}^{H}\mathbf{h}_{k} \rvert}^{2}}{\sum_{\pi_{k',i'}>\pi_{k,i}}P_{k',i'}{\lvert \mathbf{d}_{k',i'}^{H}\mathbf{h}_{k'} \rvert}^{2}+\sigma_{\textrm{ul}}^{2}}\bigg),
\end{equation}
{
\begin{equation}\label{eq:UL_Rate1}
    R_{K}=\log_{2}\bigg(1+\frac{P_{K}{\lvert \mathbf{d}_{K}^{H}\mathbf{h}_{K} \rvert}^{2}}{\sum_{\pi_{k',i'}>\pi_{K}}P_{k',i'}{\lvert \mathbf{d}_{k',i'}^{H}\mathbf{h}_{k'} \rvert}^{2}+\sigma_{\textrm{ul}}^{2}}\bigg).
\end{equation}
Since the message of user-$k,\,k\in\mathcal{K}\setminus\{K\}$ is transmitted via two streams, the achievable rate is given by $R_{k}=R_{k,1}+R_{k,2}$}. Fig.~\ref{fig:RSMA_UL} illustrates the uplink transmission framework of RSMA. From (\ref{eq:UL_Rate})-(\ref{eq:UL_Rate1}), it can be inferred that the rates of users are dependent on both the power allocation at the transmitters and decoding order at the receiver. To that end, the decoding order $\pi$ and power allocation $\{P_{k,i}, P_{K}\mid \forall k\in\mathcal{K}\setminus\{K\},\,i\in\{1,2\}\}$ are optimized to maximize the uplink sum-rate performance of RSMA in \cite{uplink@RSMA}. Uplink RSMA was shown to achieve the capacity region of the Gaussian MAC without utilizing time sharing among users\cite{uplink@MAC}. To avoid time sharing, additional complexity is introduced at the BS in the form of multiple layers of SIC. Nonetheless, RSMA significantly benefits uplink transmission and can enable core services where users have intermittent access behaviour such as URLLC and mMTC. Moreover, RSMA is capable of enabling different combinations of core services to serve users with heterogeneous profiles\cite{URLLCEmBB@Santos}.   
\section{Applications and Future Works}\label{Applications}\vspace{-0.1cm}
RSMA has attracted intensive research attention due to its appealing benefits in multi-antenna communications. In the following, we illustrate the emerging applications and potential future works of RSMA in enabling technologies and core services that are envisioned to play key roles in $6$G. 
\subsubsection{Massive antennas} Vulnerability to inaccurate CSIT hinders the capability of MaMIMO to achieve high SE, EE, and reliability. Compared to SDMA, RSMA is more robust to CSIT inaccuracies resulting from mobility, hardware impairments, pilot contamination, etc, in MaMIMO networks\cite{onur@mobility,mishra2022ratesplitting,mao2022ratesplitting}. {Similarly, RSMA is capable of benefiting other promising and potential $6$G massive antenna technologies such as reconfigurable intelligent surfaces (RIS) \cite{mao2022ratesplitting} and lens antenna arrays, which are vulnerable to CSIT inaccuracies}.
\subsubsection{Ultra high frequencies} Stringent requirements of exorbitantly high data rates have motivated the need to use ultra high frequencies for transmissions which are vulnerable to high propagation loss. MaMIMO is typically employed to offset the propagation loss which itself is sensitive to CSIT inaccuracies. RSMA achieves higher sum-rate performance and lower CSIT acquisition complexity in mmWave communications compared to SDMA \cite{mmWave@RSMA,mao2022ratesplitting}. In the future, RSMA can help to address the issues of THz communications such as inefficient user synchronization and low resolution analog-to-digital converters, in addition to legacy issues of mmWave.
\subsubsection{{Optical wireless communications (OWC)}} For high-speed short-range communications, conventional radio frequency based networks are expected to be complemented by emerging technologies such as visible light communications (VLC). RSMA achieves DoF, SE and EE gains over current MA schemes in multi-user MIMO VLC in the presence of deterring issues such as user channel correlations and imperfect CSI \cite{mao2022ratesplitting}. Other OWC technologies such as infrared and ultraviolet which exhibit similar behaviour to VLC could also benefit from RSMA in managing multi-user interference.
\subsubsection{Waveform design} Current waveform designs are incapable of handling massive heterogeneous devices in $6$G. The capabilities of RSMA can aid in flexible waveform design by addressing practical challenges such as inter-channel interference, massive asynchronous transmissions, latency due to processing delays, Doppler effect and reception under imperfections to achieve high SE and EE performance. {RSMA also has great potential of integrating with other emerging waveforms such as vortex wave\cite{chen@waveform}, etc}. Furthermore, RSMA can aid waveform design for other new enabling technologies such as {integrated sensing and communications (ISAC)}.
\subsubsection{Integration with other MA schemes} Integrating RSMA with current MA schemes can help address their respective issues such as asynchronization resulting in inter-user interference in code division MA and inter-group interference in multi-user shared access, mobility in sparse code MA, and inter-carrier interference in OFDMA. {Future MA schemes such as mode division MA can similarly utilize RSMA to deal with inter-mode interference resulting from mode offset.} 
\subsubsection{Network slicing} In $6$G, a combination of services such as eMBB and URLLC will require the network to serve heterogeneous user profiles with diverse QoS requirements. \cite{URLLCEmBB@Santos} utilizes network slicing to share the resources among users with different profiles and shows that RSMA outperforms OMA and NOMA schemes for URLLC service even in the presence of interference from eMBB users. Future works on RSMA can  investigate its role in mitigating the impact of imperfect CSIT and ensuring security in allocation of slice resources to users of different profiles. 
\subsubsection{Distributed networks} New age technologies such as blockchain have provided great impetus to distributed networks. Interference management capabilities of RSMA has the potential to address the issues of imperfect CSIT, congestion, resource allocation, etc and aid distributed networks in achieving better throughput, EE and security.
\subsubsection{mMTC}The sporadic access behaviour of the devices in mMTC complicates resource allocation which in turn deteriorates both performance and reliability. RSMA is shown to improve uplink transmission reliability in mMTC and achieve a better outage performance than NOMA\cite{liu@2021rate}. In mMTC downlink, \cite{mishra2022ratesplitting} shows that RSMA achieves SE performance gain over SDMA with deteriorated CSIT in cell free MaMIMO network. Role of RSMA in aiding the coexistence of mMTC with URLLC and eMBB is yet to be investigated.  
\subsubsection{Vehicular to everything (V$2$X) communications} RSMA can aid V$2$X communications as it is more robust to user mobility than current MA schemes\cite{onur@mobility}. RSMA also brings rate benefits over current MA schemes with shorter blocklength addressing the issue of strict latency constraints\cite{mao2022ratesplitting}. Future works on RSMA can help to address the challenges such as resource collisions in dense environments, and interference arising from spectrum sharing or network slicing between vehicular and non-vehicular users in integrated networks. 
\subsubsection{{ISAC}} {A new design paradigm for $6$G that aims at integrating sensing and communication functionalities countenances a major challenge of incurring interference between the two functionalities}. RSMA-aided dual functional radar communication (DFRC) transmission framework has been shown to improve the performance of both functionalities and reduce hardware complexity in ISAC compared to SDMA-aided DFRC \cite{Onur@ISAC,mao2022ratesplitting}. An interesting next step would be to utilize RSMA in aiding waveform design and security in ISAC.
\subsubsection{{Artificial intelligence and machine learning (ML)}} ML will play a more significant role in $6$G than current generation networks. The use of ML in RSMA-assisted network has been investigated to design power control algorithms without any prior information of the CSI and rate performance gain over SDMA-assisted network was achieved\cite{ML@RSMA}. \cite{UAV@RSMA_ML} is another instance where integration of RSMA and ML allowed better deployment of unmanned aerial vehicle (UAVs) thereby reducing transmit power levels compared to OMA in UAV-assisted networks. For future research, the only caveat to keep in mind is that the integration of RSMA and ML must not be restricted to the PHY layer only, instead the focus should be on cross-layer design such as aiding network orchestration, to truly assist wireless communication for intelligent $6$G.
\subsubsection{Intelligent edge computing} Massive number of edge users and their diverse profiles pose a significant challenge for mobile edge computing (MEC) in providing efficient resource utilization, complexity reduction and latency-critical computing in wireless networks. RSMA has been shown to aid MEC better than NOMA in offloading tasks for cognitive radio network\cite{liu2022rate}. Integrating RSMA and ML with MEC to provide intelligent edge computing will benefit tasks such as automation of network orchestration and potentially aid MEC in achieving better user fairness, performance and security.
\subsubsection{Space-air-ground integrated networks (SAGIN)} {RSMA achieves superior SE and EE performance compared to current MA schemes in satellite and UAV integrated networks\cite{mao2022ratesplitting}}. Consequently, applying RSMA for SAGIN can help dealing with formidable challenges of mobility, spectrum sharing, interoperability of different segments and heterogeneous QoS constraints. Moreover, the interplay of SAGIN with other enablers such as ML and MEC will be benefited by RSMA to achieve ubiquitous intelligent connectivity.
\subsubsection{Simultaneous wireless information and power transfer (SWIPT)} RSMA-assisted SWIPT has been shown to achieve rate performance gain over SDMA-assisted SWIPT for a given sum energy constraint of energy receivers. In addition, RSMA has been shown to save more transmit power than SDMA for co-located energy and information receivers\cite{mao2022ratesplitting}. Future works integrating RSMA and SWIPT with coverage extension techniques such as RIS can help address the challenge of limited transmission range to energy receivers and benefit services employing low-powered devices.
\section{Challenges}\label{Challenges}
A wide body of literature has been developed to demonstrate the capabilities of RSMA and establish its superiority over current MA schemes. Nevertheless, the study of RSMA is very much in its nascent stage and further research is needed to address current challenges and enhance its performance. For example, investigating alternative designs to SIC to reduce receiver complexity in downlink and devising low-complexity methods for determining message splitting ratios and power allocation to reduce signalling overhead in uplink are challenges that need attention going forward. 
\par It is also imperative to address the implementation issues that act as an impediment in exploiting the full potential of RSMA. Since RSMA splits the message of a user at the transmitter, the user needs to be aware of the knowledge of message splitting (and combining) to successfully decode and remove the common stream using SIC at the receiver. Such a requirement increases the signalling burden at the transmitter and receiver both, and should be addressed. Another implementation issue is the complexity associated with hybrid automatic repeat request (HARQ). Since HARQ is responsible for handling packet re-transmission in case the receiver detects an incorrectly decoded message, the framework of RSMA that involves message splitting and combining introduces difficulties, which are in need of being addressed. {Furthermore, PHY layer design of RSMA taking higher transmission layers (e.g., MAC layer) into account for investigating the system-level performance, and designing specific CSI feedback mechanisms for RSMA are research directions that have not been explored yet}. {Finally, it is imperative to design prototypes and testbeds to validate different solutions developed for RSMA and provide substantial impetus to its standardization\cite{mao2022ratesplitting}}.   
\section{Conclusions}\label{Conclusion}
In this letter, we overview the basic principles of the RSMA scheme and highlight its potential to be a promising NGMA scheme for future networks such as $6$G and beyond. We begin by describing the peculiar characteristics of $6$G, and the motivation behind opting RSMA as the MA for the same. We then delineate the design principle, transmission framework, applications, and future works of RSMA. In the sequence of this tutorial, Part II focuses on the interplay of RSMA with ISAC, and Part III on interplay of RSMA with RIS.

\ifCLASSOPTIONcaptionsoff
  \newpage
\fi

\appendices

\bibliographystyle{IEEEtran}
\bibliography{reference}

\end{document}